\documentclass[twocolumn,preprintnumbers,amsmath,amssymb,bibnotes,nofootinbib]{revtex4-1}

\usepackage{graphicx}
\usepackage{dcolumn}
\usepackage{bm}
\usepackage{cancel} 
\usepackage{xcolor}

\usepackage[pdf]{pstricks}
\usepackage{pst-node}
\usepackage{calc}

\usepackage{float}
\restylefloat{figure}
\usepackage{epstopdf}

\usepackage{tikz}
\usetikzlibrary{calc} \usetikzlibrary{patterns} \usetikzlibrary{decorations.pathreplacing} \usetikzlibrary{decorations.markings} \usetikzlibrary{decorations.pathmorphing} \usetikzlibrary{positioning}

\restylefloat{figure}
\usepackage{epstopdf}

\newcommand{\be}{\begin{equation}}
\newcommand{\ee}{\end{equation}}
\newcommand{\comment}[1]{}

\newcommand{\nwc}{\newcommand}
\nwc{\ba}  {\begin{array}}
\nwc{\ea}  {\end{array}}
\nwc{\bdm} {\begin{displaymath}}
\nwc{\edm} {\end{displaymath}}

\nwc{\bea} {\begin{equation}\ba{lcl}}
\nwc{\eea} {\ea\end{equation}}

\nwc{\bda} {\bdm\ba{lcl}} 
\nwc{\eda} {\ea\edm}

\nwc{\bc}  {\begin{center}}
\nwc{\ec}  {\end{center}}

\def\eqn#1{eq.~(\ref{#1})}

\nwc{\ds}  {\displaystyle}
\nwc{\bmat}{\left(\ba}
\nwc{\emat}{\ea\right)}
\nwc{\nn}  {\nonumber}
\nwc{\nnn} {\nonumber \vspace{.2cm} \\ }
\nwc{\ra}  {\rightarrow}
\nwc{\lra} {\longrightarrow}

\nwc{\p} {\partial}

\def\beq{\begin{equation}}
\def\eeq{\end{equation}}

\newcommand{\vecb}{\left(\begin{array}{c}}
\newcommand{\vece}{\end{array}\right)}
\newcommand{\ccb}{\left(\begin{array}{cc}}
\newcommand{\cce}{\end{array}\right)}
\newcommand{\cccb}{\left(\begin{array}{ccc}}
\newcommand{\ccce}{\end{array}\right)}
\newcommand{\ccccb}{\left(\begin{array}{cccc}}
\newcommand{\cccce}{\end{array}\right)}
\newcommand{\cccccb}{\left(\begin{array}{ccccc}}
\newcommand{\ccccce}{\end{array}\right)}

\newcommand{\si}{\sigma}

\newcommand{\om}{\omega}

\newcommand{\Ga}{\Gamma}

\newcommand{\te}{\textrm}

\newcommand{\dd}{\mathrm{d}}

\newcommand{\RR}{\mathbb R}

\newcommand{\CC}{\mathbb C}
\newcommand{\NN}{\mathbb N}

\begin{document}

\preprint{AEI--2013--195, DAMTP--2013--23, MPP--2013--120}

\title{All order $\bm{\alpha'}$-expansion of superstring trees from the Drinfeld associator}

\author{Johannes Broedel$^a$,
Oliver Schlotterer$^{b,c}$,
Stephan Stieberger$^d$ and Tomohide Terasoma$^e$}
\affiliation{$^a$ Institut f\"ur theoretische Physik, Eidgen\"ossische Technische Hochschule Z\"urich, 8093 Z\"urich, Switzerland}
\affiliation{$^b$ Max-Planck-Institut f\"ur Gravitationsphysik, Albert-Einstein-Institut,
14476 Potsdam, Germany,}
\affiliation{$^c$ Department of Applied Mathematics and Theoretical Physics, Cambridge CB3 0WA, United Kingdom,}
\affiliation{$^d$ Max-Planck-Institut f\"ur Physik, Werner-Heisenberg-Institut,
80805 M\"unchen, Germany}
\affiliation{$^e$ Department of Mathematical Science, University of Tokyo, Komaba 3-8-1, Meguro, Tokyo 153, Japan.}

\begin{abstract}
  We derive a recursive formula for the $\alpha'$-expansion of superstring tree
  amplitudes involving any number $N$ of massless open string states. String
  corrections to Yang-Mills field theory are shown to enter through the
  Drinfeld associator, a generating series for multiple zeta values. Our
  results apply to any number of spacetime
  dimensions or supersymmetries and chosen helicity configurations.
\end{abstract}

\maketitle

\section{Introduction}

Scattering amplitudes are the most fundamental observables in both quantum
field theory and string theory. In recent years, numerous hidden structures
underlying the S-matrix have been revealed in both disciplines. Several of
these discoveries can be attributed to and have benefited from the close
interplay between amplitudes of string theory in the low-energy limit and
supersymmetric Yang-Mills (YM) field theory. 

A main challenge in the study of field theory amplitudes originates from the
transcendental functions in their quantum corrections. Novel mathematical
techniques such as the symbol \cite{GoncharovOrig} helped to streamline the
polylogarithms and multiple zeta values (MZVs) in loop amplitudes of (super-)YM
theory. In string theory, MZVs appear in the $\alpha'$-corrections already at
tree level due to the exchange of infinitely many heavy vibrational modes.
These effects are encoded in integrals over world-sheets of genus zero.

The study of $\alpha'$-expansions in the superstring tree-level amplitude is
interesting from both a mathematical and a physical point of view. On the one
hand, the pattern of MZVs appearing therein can be understood from an
underlying Hopf algebra structure \cite{Schlotterer:2012ny}.  On the other
hand, explicit knowledge of the associated string corrections is crucial for
the classification of candidate counterterms in field theories with unsettled
questions about their UV properties \cite{Beisert:2010jx}. 

In spite of technical advances to evaluate $\alpha'$-expansions for any
multiplicity \cite{polylog}, compact and straightforwardly applicable formulae
for string corrections are still lacking. This letter closes this gap by
describing a novel method to recursively determine the $\alpha'$-dependence of
$N$-point trees through the generating function of MZVs -- the Drinfeld
associator. Its connection with superstring amplitudes -- in particular the
common pattern of MZV appearance -- was firstly pointed out in
\cite{Drummond:2013vz}. Our techniques are based on the Knizhnik-Zamolodchikov
(KZ) equation \cite{Knizhnik:1984nr} obeyed by world-sheet integrals and
thereby resemble ideas in field theory to determine loop integrals
\cite{Henn:2013pwa}. Along the lines of \cite{Terasoma}, the associator is
shown to connect boundary values, given by $N$-point and $(N-1)$-point disk
amplitudes, respectively. The method presented in this article bypasses the
cumbersome direct evaluation of world-sheet integrals and reduces their
$\alpha'$-expansions to simple matrix multiplications. Apart from its
conceptual accessibility, it substantially reduces the computational effort in
deriving the explicit form\footnote{The website \cite{www}
provides
expressions for string corrections to five- to seven-point
amplitudes as well as material to apply the presented method up to nine-points.}
of $\alpha'$-corrections.

\medskip
{\bf A. The structure of disk amplitudes:} 
The color-ordered $N$-point disk amplitude $A_{\te{open}}(\alpha') := 
A_{\te{open}}(1,2,\ldots,N;\alpha')$ was computed in
\cite{Mafra:2011nv,Mafra:2011nw} based on pure spinor cohomology methods
\cite{Mafra:2010jq}. Its entire polarization dependence was found to enter
through color-ordered tree amplitudes $A_{\te{YM}}$ of the underlying
YM field theory which emerges in the point particle limit $\alpha'
\rightarrow 0$:
\beq
A_{\te{open}}(\alpha') = \sum_{\si \in S_{N-3}} F^\si(\alpha') \, A_{\te{YM}}^\si\,.
\label{tree}
\eeq
The $(N-3)!$ linearly independent \cite{BCJ} subamplitudes\footnote{Labels
  $1,2,\ldots,N$ in the subamplitude eq. (\ref{tree})
denote any state in the gauge supermultiplet.}
$A_{\te{YM}}(1,\si(2,3,\ldots,N-2),N-1,N)$ are grouped into a vector
$A_{\te{YM}}^\si$.
The objects $F^\si(\alpha')$ describe string corrections to YM amplitudes and
will be recursively determined as the main result of this letter. They are
generalized Selberg integrals \cite{Selberg} over the boundary of the open
string world-sheet of disk topology:
\begin{align}
F^\si& = (-1)^{N-3}\prod_{i=2}^{N-2}\int_{z_{i}< z_{i+1}} \! \! \! \! \! \dd z_i \ {\cal I} \, \si \bigg\{ \prod_{k=2}^{N-2} \sum_{j=1}^{k-1} \frac{s_{jk}}{z_{jk}} \bigg\}
\label{Fs}\, , \\
{\cal I} &= \prod_{i<j}^{N-1} |z_{ij}|^{s_{ij}} \ , \ \ \ (z_1,z_{N-1},z_N) = (0,1,\infty)\,. 
\label{choice}
\end{align}
The $S_{N-3}$ permutation $\si$ acts on labels $2,3,\ldots,N\!-\!2$ of $z_{ij}
:= z_i\!-\!z_j$ and of the dimensionless Mandelstam invariants
\beq
s_{i_1 i_2 \ldots i_p} = \alpha' (k_{i_1}+k_{i_2}+\ldots +k_{i_p})^2 \ ,
\label{mand}
\eeq
which carry the $\alpha'$-dependence of the string amplitude
(\ref{tree}). 
The $k_i$ denote external on-shell momenta. 
Hence, the
$s_{ij}$-expansion of the integrals (\ref{Fs}) encodes the low energy behaviour
of superstring tree amplitudes.



\medskip
{\bf B. Multiple zeta values:} 
As discussed in both mathematics \cite{Aomoto, Terasoma, Francis} and physics
\cite{Stieberger:2009rr, Mafra:2011nw, Schlotterer:2012ny} literature, the
$\alpha'$-expansion of Selberg integrals involves (products of) MZVs.
They can be defined by iterated integrals over differential forms $\om_0 := \frac{\dd z}{z}$ and $\om_1 :=
\frac{\dd z}{1-z}$
\beq
\zeta_{n_1,\ldots,n_r} = \!\!\!\!\!\! \!\!\!\!\!\!  \int\limits_{0<z_i<z_{i+1}<1} \!\!\!\!\!\!  \!\!\!\!\!\! \om_1 \underbrace{\om_0 \ldots \om_0}_{n_1-1} \om_1 \underbrace{\om_0 \ldots \om_0}_{n_2-1} \ldots \om_1 \underbrace{\om_0 \ldots \om_0}_{n_r-1}
\label{mzvdef}
\eeq
where $n_j \in \NN$ and $n_r \geq 2$. The overall weights $\sum_{j=1}^r n_j$ of MZV factors match the power of $\alpha'$ in the string amplitudes' expansion. Instead of labeling MZVs by the set of $n_j$, one can equivalently encode the
integrand of eq.~(\ref{mzvdef}) in a word $w$ in the alphabet $\{0,1\}$ (i.e.~$w \in \{ 0,1\}^{\times}$) where the function $w[\om_0,\om_1]$ translates this word into sequences of
$\{\om_0,\om_1\}$ \cite{Drummond:2013vz}:
\beq
\zeta_{ ( w ) } := 
\!\!\!\!\!\! \!\!\!\!\!\!  \int\limits_{0<z_i<z_{i+1}<1} \!\!\!\!\!\!  \!\!\!\!\!\! w[\om_0,\om_1]\,.
\label{notation}
\eeq
The pattern of MZVs in the $\alpha'$-expansion of (\ref{Fs}) has been revealed
in \cite{Schlotterer:2012ny} on the basis of a Hopf algebra structure.

\medskip
{\bf C. The Drinfeld associator:} 
Consider the KZ equation with $z_0 \in \CC\backslash\{0,1\}$ and Lie-algebra generators $e_0,e_1$:
\beq
\frac{\dd {\bf{\hat F}}(z_0)}{\dd z_0} = \left( \frac{ e_0 }{z_0} + \frac{ e_1}{1-z_0 } \right) {\bf{\hat F}} (z_0) \ .
\label{KZ}
\eeq
The solution ${\bf{\hat F}}(z_0)$ of the KZ equation takes values in the vector
space the representation of $e_0$ and $e_1$ is acting upon.  The regularized
boundary values
\beq
C_0 := \lim_{z_0 \rightarrow 0} z_0^{-e_0} {\bf{\hat F}}(z_0) \ , \ \ \ C_1 := \lim_{z_0 \rightarrow 1} (1-z_0)^{e_1} {\bf{\hat F}}(z_0)\,
\label{C01}
\eeq
are related by the Drinfeld associator \cite{Drinfeld:1989st, Drinfeld2}
\beq
C_1 = \Phi(e_0,e_1)\,C_0 \ ,
\label{C0toC1}
\eeq
where $C_0$, $C_1$ and $\Phi$ take values in the universal enveloping algebra
of the Lie algebra generated by $e_0$ and $e_1$.  The regularizing factors
$z_0^{-e_0}$ and $(1-z_0)^{e_1}$ are included into \eqn{C01} as to render the
$z_0\rightarrow 0,1$ regime of ${\bf{\hat F}}(z_0)$ real-single-valued. In the
notation of \eqn{notation}, the Drinfeld associator can be represented as a
generating series of MZVs \cite{Le}:
\beq
\Phi(e_0,e_1) = \sum_{w \in \{0,1\}^{\times}}  \tilde w[e_0,e_1] \zeta_{ ( w ) }\,,
\label{drin}
\eeq
where $\tilde{w}$ denotes the reversal of the word $w$. The series expansion
of \eqn{drin} in a basis of MZVs starts with the following commutators $[\cdot,\cdot ]$:
\begin{align}
\Phi(e_0,e_1)& = 1 + \zeta_2 [e_0,e_1] + \zeta_3 [e_0-e_1,[e_0,e_1]] \notag \\
 &+ \zeta_4 ([e_0,[e_0,[e_0,e_1]]]+\tfrac{1}{4}[e_1,[e_0,[e_1,e_0]]] \notag \\
 & \ \ -[e_1,[e_1,[e_1,e_0]]]+ \tfrac{5}{4} [e_0,e_1]^2)+  \ldots\;\; \ ,
\end{align}


\medskip
{\bf D. Main result:} 
In this letter, we identify the Drinfeld associator $\Phi$ as the link between
$N$-point string amplitudes and those of multiplicity $N-1$. Thus, starting
from the $\alpha'$-independent three-point level, one can build up any
tree-level string amplitude recursively. 

We will construct a matrix representation for the associator arguments $e_0$
and $e_1$ in section I.C for each multiplicity. Starting with a boundary value $C_0$
containing the world-sheet integrals for the $(N\!-\!1)$-point amplitude,
\eqn{C0toC1} yields a vector $C_1$, which we will show to encode the integrals
\eqn{Fs} for multiplicity $N$.  Consequently, one can express the $N$-point
world-sheet integrals $F^\si$ in terms of those at $(N\!-\!1)$-points
\begin{align}
F^{\si_i} &= \! \sum_{j=1}^{(N-3)!} \! \big[ \Phi(e_0,e_1) \big]_{ij} 
\ F^{\si_j } \big|_{k_{N-1} =0},
\label{main} 
\end{align}
where the soft limit $k_{N-1}=0$ gives rise to $(N-1)$-point integrals on the
right hand side
\begin{align}
  F^{\si(23\ldots N-2)} \big|_{k_{N-1} =0}& = \left\{ \begin{array}{cl} \!F^{\si(23\ldots N-3)}\! &\te{if}\,\,\si(N\!-\!2)\!=\!N\!-\!2 \\
0\!\! & \te{otherwise\,.} \end{array} \right.
\label{rec}
\end{align}
The permutations $\si_i$ are canonically ordered in \eqn{main}.

\section{The method}
\label{sec:method}

The backbone of the recursion \eqn{main} is a vector $\bf{\hat F}$ of auxiliary
functions and a corresponding matrix representation of $e_0,e_1$ such that the
KZ equation (\ref{KZ}) holds 
. Moreover, the boundary values $C_0$ and $C_1$ derived from
$\bf{\hat F}$ via \eqn{C01} need to reproduce basis functions \eqn{Fs} of
multiplicity $N-1$ and $N$, respectively. As we will see, these requirements
are met by components
\begin{align}
  \hat F_\nu^\si&(z_0,s_{0k}) = (-1)^{N-3} \int^{z_0}_0 \dd z_{N-2} \prod_{i=2}^{N-3}\int_0^{z_{i+1}} \dd z_i \ {\cal I}  \notag \\  
   & \times \prod_{k=2}^{N-2} z_{0k}^{s_{0k}} \si \bigg\{\prod_{k=2}^{\nu} \sum_{j=1}^{k-1} \frac{ s_{jk}}{z_{jk}} \prod_{m=\nu+1}^{N-2} \sum_{n=m+1}^{N-1} \frac{ s_{mn}}{z_{mn}}\bigg\} \,.
\label{Faux}
\end{align}
The vector $\bf{\hat{F}}$ is composed from $N\!-\!2$ subvectors $\hat{F}_\nu$
of length $(N\!-\!3)!$. Numbered by $\nu\!=\!1,2,..,N\!-\!2$, they appear in
decreasing order, that is, ${\bf{\hat
F}}=(\hat{F}_{N-2},\hat{F}_{N-3},\ldots,\hat{F}_1)$. Entries of the
$\hat{F}_\nu$ are labeled by permutations $\si\in S_{N-3}$.


\begin{figure}[hb]
    \includegraphics{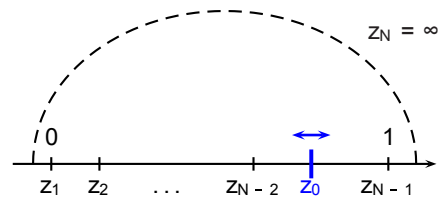}
\caption{Worldsheet with an auxiliary position $z_0$.}
\end{figure}

The integrals in \eqn{Faux} generalize the functions \eqn{Fs} through an
auxiliary world-sheet position $z_0$ and
auxiliary Mandelstam variables $s_{0k}$\footnote{As will be explained below, we
  will eventually set $s_{0k} \rightarrow 0$ and therefore do not display them
  as arguments of ${\bf \hat{F}}(z_0)$.}. 
This $z_0$ enters in the integration
limit of the outermost integral as well as in the deformation
$\prod_{k=2}^{N-2} (z_{0k})^{s_{0k}}$ of the Koba-Nielsen factor ${\cal I}$
and serves as the differentiation variable for the KZ equation (\ref{KZ}). As visualized in the above figure, the position $z_0$ downscales the integration domain on the disk boundary and thus interpolates between world-sheet-configurations of an $N$-point and $(N\!-\!1)$-point tree amplitude.

 At $z_0\!=\!1$ and $s_{0k}\!=\!0$ -- in absence of the augmentation --
 the functions $\hat F_\nu^\si$ in \eqn{Faux} approach the integrals 
$F^\si$ in the amplitude for any $\nu$. 
In this regime, $\nu$ labels different equivalent
representations \cite{polylog} of the integrals \eqn{Fs}. 

Matching the length of the auxiliary vector, $e_0$ and $e_1$ in \eqn{KZ} are $(N\!-\!2)!\times(N\!-\!2)!$-matrices. It is
known \cite{Terasoma} that their entries are linear forms on $s_{ij}$. They can
be determined by matching the $z_0$ derivatives\footnote{The boundary term from
  acting with $\frac{ \dd }{\dd z_0}$ on the integration limit does not
  contribute as can be seen by analytic continuation of
  $(z_{0,N-2})^{s_{0,N-2}} |_{z_{N-2}=z_0} = 0 \ \forall
\ s_{0,N-2} \in \RR^+$.} of $\hat F^\si_\nu$ with the
right hand side of the KZ equation (\ref{KZ}). Once the resulting matrices
$e_0$ and $e_1$ are available, one can calculate the Drinfeld associator to any
desired order employing its series expansion \eqn{drin}. Having set up the KZ
equation (\ref{KZ}) for the auxiliary function $\bf{\hat{F}}$, we will now
relate its regularized boundary terms \eqn{C01} to the integrals \eqn{Fs} in
the string amplitude.

\medskip
{\bf A. The $\bm{z_0 \rightarrow 0}$ boundary value $\bm{C_0}$:} 
The boundary term $C_0$ is determined by taking the limit $z_0\! \rightarrow
\!0$ of $z_{0}^{-e_0}{\bf{\hat F}} (z_0)$. This amounts to squeezing the
world-sheet positions $z_2,\ldots,z_{N-2}$ into an interval $[0,z_0]$ of
vanishing size, see the above figure. This effectively removes 
one of the
$N-3$ integrations and makes contact with the $(N-1)$-point problem. Let us
make this more precise:
The first $(N\!-\!3)!$ components of ${\bf{\hat
F}}(z_0\! \rightarrow \! 0)$ at $\nu=N\!-\!2$,
\beq
\hat F_{N-2}^\si(z_0 \rightarrow 0,s_{0i}) = z_0^{s_{\te{max}}} F^\si\big|_{ s_{i,N-1} =  s_{0i}}+ {\cal O}(s_{0i})\,,
\label{simpl0}
\eeq
involve the eigenvalue $s_{\te{max}}\! =\! s_{12\ldots N-2} \!+\!
\sum_{j=2}^{N-2} s_{0j}$ of $e_0$ \cite{Terasoma}. The remaining subvectors of
${\bf{\hat F}}(z_0\!\rightarrow\!0)$ at $\nu \leq N\!-\!3$ are suppressed by
$N-2-\nu$ powers of $z_0$\footnote{This can be seen by a change of integration
  variables $z_i = z_0 w_i$ rescaling the integration region to $0 \!\leq\! w_2
  \! \leq \! w_3 \! \leq \! \ldots \! \leq \! w_{N-2} \! \leq \! 1$.} and do
  not contribute to $C_0$.  The action of $z_0^{-e_0}$ compensates the $z_{0}$
  dependence of the resulting vector $(z_{0}^{s_{\te{max}}} F^{\si}, {\bf
  0}_{(N-3) (N-3)!})$. 

The desired $(N-1)$-point integrals can be achieved through a soft limit $k_{N-1} \rightarrow 0$,
see (\ref{rec}). This can be realized by setting $s_{0i}=s_{i,N-1}=0$ in
\eqn{simpl0} which converts the subvector $\hat F^{\sigma}_{N-2}$ into
$(N-1)$-point data
\beq
C_{0}  = ( F^\si\big|_{ k_{N-1} =  0} , {\bf 0}_{(N-3) (N-3)!}) \ .
\label{NptC00}
\eeq

\medskip
{\bf B. The $\bm{z_0 \rightarrow 1}$ boundary value $\bm{C_1}$:} The $z_0
\rightarrow 1$ regime of $(1-z_{0})^{e_1} {\bf{\hat F}}(z_0)$ underlying $C_1$
restores the integration domain of the $N$-point functions \eqn{Fs}.
Considering the schematic form of the first $(N\!-\!3)!$ rows in
\beq
(1-z_{0})^{e_1} = \left(\begin{array}{cc} \bm{1}_{(N\!-\!3)!\times (N\!-\!3)!} &{\bf 0}_{(N\!-\!3)!\times (N\!-\!3) (N\!-\!3)!}  \\ \vdots&\vdots \end{array} \right) 
\label{block}
\eeq
we can neglect all components of ${\bf{\hat F}}(z_0 \rightarrow  1)$ except 
\beq
\hat F_{N-2}^\si(z_0 \rightarrow 1,s_{0i}) = F^\si + {\cal O}(s_{0i}) \ .
\label{simpl1}
\eeq
Setting $s_{0i}=0$ as motivated in section II.A leads
to
\beq
C_1 = (F^{\si}, \ldots) \ .
\label{NptC1}
\eeq
Our setup does not require the delicate evaluation of the
remaining 
components in the ellipsis.

\medskip
{\bf C. Summary:} Our main result \eqn{main} follows by specializing the
central property \eqn{C0toC1} of the associator to the representations of
$C_i,e_i$ extracted from the auxiliary vector ${\bf{\hat F}}(z_0)$ defined in
\eqn{Faux}. In \eqn{NptC00} and \eqn{NptC1}, we have identified $C_0$ and $C_1$
with $(N-1)$- and $N$-point world-sheet integrals \eqn{Fs}, respectively. This
turns \eqn{C0toC1} into a recursion in $N$ where the arguments $e_0,e_1$ of the
connecting associator can be straightforwardly read off from the KZ equation
(\ref{KZ}) satisfied by ${\bf{\hat F}}(z_0)$. Starting from the trivial
three-point amplitude, this allows to determine the complete
$\alpha'$-expansion to any order and for any multiplicity.  

\section{Examples}
\label{sec:expl}

\medskip
{\bf A. From $\bm{N=3}$ to $\bm{N=4}$:} 
Any four--point disk integral is proportional to
\[
F^{(2)}\!=\!  \int^1_0 \!\!\dd z_2 \, |z_{12}|^{s_{12}} |z_{23}|^{s_{23}} \, \frac{s_{12}}{z_{21}} = \frac{ \Ga(1+s_{12}) \Ga(1+s_{23}) }{\Ga(1+s_{12}+s_{23})}\,.
\]
We will rederive its $\alpha'$-expansion from the Drinfeld associator along the
lines of section \ref{sec:method}. The auxiliary vector \eqn{Faux} contains two
subvectors of length one:
\beq
\vecb \hat F^{(2)}_2 \\ \hat F^{(2)}_1 \vece =  \int^{z_0}_0 \dd z_2 \, |z_{12}|^{s_{12}} |z_{23}|^{s_{23}} z_{02}^{s_{02}}\, \vecb s_{12}/z_{21} \\ s_{23}/z_{32} \vece \ .
\label{aux4}
\eeq
Partial fraction decomposition $(z_{12} z_{02})^{-1} = (z_{12}
z_{01})^{-1}-(z_{01} z_{02})^{-1}$ followed by discarding a $z_2$-derivative
\beq
0 = \int \dd z_2 \, |z_{12}|^{s_{12}} |z_{23}|^{s_{23}} z_{02}^{s_{02}} \left( \frac{s_{02}}{z_{02}}+\frac{s_{12}}{z_{12}}-\frac{s_{23}}{z_{23}} \right)
\label{IBP}
\eeq
leads to the following KZ equation after setting $s_{02}=0$:
\begin{align}
& \ \ \frac{ \dd }{ \dd z_0 } \vecb \hat F^{(2)}_2 \\ \hat F^{(2)}_1 \vece = \left( \frac{ e_0 }{z_{01}} - \frac{ e_1 }{z_{03}} \right) \vecb \hat F^{(2)}_2 \\ \hat F^{(2)}_1 \vece \ ,
\label{KZ4} \\
&e_0 = \ccb s_{12} &-s_{12} \\ 0 &0 \cce \ , \ \ \ e_1 = \ccb 0 &0 \\ s_{23} &-s_{23} \cce \ .
\label{e4}
\end{align}
The regularized boundary values (\ref{C01}) read
\beq
C_0 = \vecb 1 \\ 0 \vece \ , \ \ \ C_1 = \vecb F^{(2)} \\ \ldots \vece 
\label{C4}
\eeq
and \eqn{main} becomes
\beq
 \vecb F^{(2)} \\ \ldots \vece = \big[ \Phi(e_0,e_1) \big]_{2 \times 2}   \vecb 1 \\ 0 \vece
\eeq
with $e_0,e_1$ given in \eqn{e4}. Their particular form implies that products
of any two matrices $\te{ad}_{0}^k \te{ad}_{1}^l [e_0,e_1]$ with $k,l \in
\NN_0$ vanish, where $\te{ad}_{i}x := [e_i,x]$. According to
\cite{Drummond:2013vz}, this allows to express the four-point disk amplitude
exclusively in terms of single $\zeta$'s ($r=1$ in \eqn{mzvdef}).

\medskip
{\bf B. From $\bm{N=4}$ to $\bm{N=5}$:} 
Next we shall derive a closed formula expression for the five-point versions
$F^{(23)}$ and $F^{(32)}$ of \eqn{Fs} by applying the associator method to the
auxiliary functions \eqn{Faux} at $N=5$
\[
\vecb  \hat F^{(23)}_3 \\ \hat F^{(32)}_3 \\ \hat F^{(23)}_2 \\ \hat F^{(32)}_2 \\ \hat F^{(23)}_1 \\ \hat F^{(32)}_1 \vece = \int \limits^{z_0}_0  \! \dd z_3 \int \limits^{z_{3}}_0 \! \dd z_2 \, {\cal I} \, z_{02}^{s_{02}} z_{03}^{s_{03}} \!  \vecb X_{12}(X_{13}\! + \!X_{23}) \\ X_{13} (X_{12}\!+\!X_{32}) \\ X_{12} X_{34} \\ X_{13} X_{24} \\ (X_{23}\!+\!X_{24}) X_{34} \\ (X_{32}\!+ \!X_{34}) X_{24} \vece
\label{aux5}
\]
where $X_{ij} := \frac{s_{ij}}{ z_{ij}}$. Partial fraction and integration by
parts analogous to (\ref{IBP}) leads to the $(6\times 6)$-matrices
\begin{align*}
e_0 = \left( \begin{array}{cccccc}
s_{123} &0 &-s_{13}-s_{23} &-s_{12} &-s_{12} &s_{12} \\
0 &s_{123} &-s_{13} &-s_{12}-s_{23} &s_{13} &-s_{13} \\
0 &0 &s_{12} &0 &-s_{12} &0 \\
0 &0 &0 &s_{13} &0 &-s_{13} \\
0&0&0&0&0&0 \\
0&0&0&0&0&0
\end{array} \right) \\
e_1 = \left( \begin{array}{cccccc}
0&0&0&0&0&0 \\
0&0&0&0&0&0 \\
s_{34} &0 &-s_{34} &0 &0 &0 \\
0 &s_{24} &0 &-s_{24} &0 &0 \\
s_{34} &-s_{34} &s_{23}+s_{24} &s_{34} &-s_{234} &0 \\
-s_{24} &s_{24} &s_{24} &s_{23}+s_{34} &0 &-s_{234} \end{array} \right)
\end{align*}
for which the KZ equation (\ref{KZ}) is satisfied after setting
$s_{02}=s_{03}=0$. The corresponding $(6 \times 6)$ associator connects the
boundary values $C_0$ and $C_1$
\beq
C_0 = \vecb F^{(2)} \\ 0 \\ {\bf 0}_4 \vece \ , \ \ \ C_1 = \vecb F^{(23)} \\ F^{(32)} \\ \vdots \vece
\eeq
via \eqn{C0toC1}, i.e.$\!$ we recursively obtain the desired $F^{(23)}$ and $F^{(32)}$ from
\beq
\vecb F^{(23)} \\ F^{(32)} \\ \vdots \vece = \big[ \Phi(e_0,e_1) \big]_{6 \times 6}   \vecb F^{(2)} \\ 0 \\ {\bf 0}_4 \vece  \ .
\label{F2332}
\eeq
Given that the four-point amplitude $\sim F^{(2)}$ only involves simple zeta
values $\zeta_n$, all the MZVs (\ref{mzvdef}) of depth $r \geq 2$ occurring in
the five-point integrals $F^{(23)}$ and $F^{(32)}$ (see
\cite{Schlotterer:2012ny} for their appearance at weights $w \leq 16$) emerge
from the associator in \eqn{F2332}.

\medskip
{\bf C. Higher multiplicity:}
The techniques to simplify derivatives of $ {\bf{\hat F}}(z_0)$ and to identify
the matrices $e_0,e_1$ in the KZ equation (\ref{KZ}) are universal to all
multiplicities. Expressions for $e_0,e_1$ up to nine points are provided at
\cite{www}, and the resulting $\alpha'$-corrections at $N=8,9$ have been
unknown 
before. Higher $N$-representations of $e_0,e_1$ are not only straightforward to compute but also
 suggested by the explicit form of their lower multiplicity cousins.
The efficiency of the associator-based recursion \eqn{main} becomes
particularly apparent at large multiplicities: The straightforward derivation of $e_0,
e_1$ avoids the growing manual effort (such as pole treatment) required by the
method of \cite{polylog}.

\section{Conclusions and outlook}

In our main result, \eqn{main}, we relate the world-sheet integrals \eqn{Fs}
carrying the $\alpha'$-dependence of $N$-point disk amplitudes to
$(N\!-\!1)$-point results by the Drinfeld associator $\Phi(e_0,e_1)$. The
challenge of evaluating world-sheet integrals is converted to elementary matrix
multiplications among $N$-dependent representations of $e_0,e_1$.

The construction works for any multiplicity and -- in principle -- to any order
in $\alpha'$. It produces previously inaccessible results, e.g.~through the
explicit form of $e_0,e_1$ for $N \leq 9$ available from \cite{www}. At
lowest orders in $\alpha'$, the new results at $N=8,9$ have been checked to
preserve the amplitudes' collinear limits, cyclicity and monodromy relations \cite{BjerrumBohr:2009rd, Stieberger:2009hq}.

The different origin of $\alpha'$-corrections therein from either the
associator or the lower point integrals might shed light on the
arrangement of reducible and irreducible diagrams in the underlying low
energy effective action \cite{prog}.

The string corrections are universal to massless open superstring tree
amplitudes in any number of spacetime dimensions, independent on the amount of
supersymmetry or chosen helicity configurations. Their $\alpha'$-expansion in
terms of MZVs can be directly carried over to closed string trees which are
expressed in terms of a specific subsector of the open string's expansion
\cite{Schlotterer:2012ny}. It would be desirable to extend this analysis to
higher genus such as the maximally supersymmetric one loop amplitudes
calculated in~\cite{Mafra:2012kh}.

\vskip0.5cm
{\bf Acknowledgments:}
We are grateful to the organizers (in particular Herbert Gangl) of the workshop
``Grothendieck-Teichmueller Groups, Deformation and Operads'' held at the
Newton Institute from January until April 2013 for creating a fruitful
framework to initiate this work. We would like to thank Claude Duhr, Herbert
Gangl and Carlos Mafra for stimulating discussions and helpful comments on the
draft. The work of OS is supported by Michael Green and the European Research
Council Advanced Grant No. 247252.

\nocite{*}
\bibliography{Drinfeld}
\bibliographystyle{h-physrev5}

\end{document}